\documentclass[preprint,12pt]{elsarticle}

\usepackage{amssymb}
\usepackage{amsmath} 
\usepackage{xcolor}
\newcommand\red[1]{\textcolor{red}{#1}} 
\newcommand\blue[1]{\textcolor{blue}{#1}} 

\begin{document}

\begin{frontmatter}



\title{Rigidity paradox of kirigami arches}


\author[inst1]{Eszter Fehér}

\affiliation[inst1,inst2]{organization={Department of Morphology and Geometric Modeling},
            addressline={Muegyetem rkp. 1-3}, 
            city={Budapest},
            postcode={1111}, 
            country={Hungary}}

\affiliation[inst2]{organization={HUN-REN-BME Morphodynamics Research Group},
            addressline={Muegyetem rkp. 1-3}, 
            city={Budapest},
            postcode={1111}, 
            country={Hungary}}

\begin{abstract}

The geometry of bending-active kirigami arches, decorated by cuts and holes, is strongly influenced by the location and geometry of the perforations. This study demonstrates that, in some instances, the geometric stiffening induced by additional cuts can outweigh the weakening effect of material removal, leading to a counterintuitive increase in structural rigidity under a given concentrated load. We present multiple parametric cut patterns to show that rigidity can be increased both under symmetric and asymmetric loads. While the preferred cut location is often near the point of action of the load, asymmetric loading can shift this optimum elsewhere. Moreover, the distance between the supports also plays a crucial role, namely, the rigidity gain vanishes when the supports are too far apart. We found that the rigidity can be increased for both non-perforated and perforated sheets, and there is a non-monotonic relationship between the global porosity and the rigidity of the structure. Numerical predictions are validated against experimental measurements.

\end{abstract}




\end{frontmatter}


\section{Introduction}
\label{sec:introduction}

The Japanese paper technique kirigami gained significant attention in the last few years due to its versatility and power in tailoring both the geometrical and mechanical properties of thin sheets. By introducing cuts and holes, kirigami could remove internal constraints of the structure, leading to an increased degree of kinematic freedom. As a result, the initially flat sheet could be transformed into three-dimensional shapes by applying external loads. The strength of the technique is that the cut pattern, the boundary conditions and the external loads determine the final shape, meaning that the structure "automatically" takes the designed shape. This unique capability has inspired applications in diverse fields, including soft robotics \cite{Hong2022}, mechanical metamaterials \cite{Hamzehei2024}, architecture \cite{Montalvo2024}, kinetic and deployable structures \cite{Lee2022, bi2023design}, biomedical devices \cite{Wu2025}, space structures \cite{Pedivellano2024}, flexible electronics \cite{Liu2022}. 

There are multiple forms of kirigami, depending on the cut pattern and the applied external load. One of the simplest kirigami structures consists of thin, initially flat strips that are usually generated by parallel cuts. These strips can bend out of the plane when their ends are brought closer to each other, which is allowed by the additional degrees of freedom. These bent strips can have further perforations and cuts that can alter their shape. In this work, we refer to these strips as \emph{kirigami arches}, emphasizing that they might have uneven edges and internal holes. If the strips are arranged sequentially, they form a cylindrical surface; if they are radially placed, they form a dome-like surface. Since the final geometry depends strongly on the location and size of the cuts, inverse design methods were developed to determine cut patterns of a desired shape \cite{Liu2020, Zhang2022, Ye2024, Rodriguez2022}.

However, cuts influence not only the shape but also the mechanical properties of the sheet \cite{Tao2023}. While cuts generally weaken the sheet, they can also redistribute the stresses and introduce geometric stiffening. There are multiple examples for improving mechanical properties of thin sheets: stability properties \cite{Blesa2019,Eisentrager2022,Tang2017,Sadik2021}, Poisson's ratio \cite{Grima2000, Grima2006, Du2024}, stretchability \cite{Cho2014,Zheng2022,Dijvejin2020}, dynamic properties \cite{Zhu2018,Khosravi2022,Li2020,Li2021}, bending properties \cite{Shrimali2021}. For a kirigami arch, an additional cut might allow for larger curvatures, leading to higher structural shapes or altered symmetry properties, both of which can affect the structure's mechanical properties. As a result, there is an interplay between the weakening effect of the cuts and the strengthening effect of the new shape. 

Zhang et al. \cite{Zhang2022} analyzed the structural rigidity of domes created from perforated sheets. In their work, for the investigated patterns, the structural rigidity and global porosity were inversely correlated. In contrast, Fehér and Gyetvai \cite{Feher2024,Gyetvai2023} presented cases where larger global porosity corresponded to higher rigidity; while Gyetvai \cite{Gyetvai2023} presented experiments in which increasing the global porosity of a given cut pattern did not affect or slightly increased the rigidity of the structure. These results suggest the possibility of optimizing cut patterns of kirigami arches for given loads and improve mechanical properties of existing cut patterns by additional cuts. 

This work aims to carry out a parametric analysis of kirigami cuts to point on the rigidity paradox of kirigami arches: additional cuts can increase the structural rigidity. We focus on a single strip, restricting the analysis to small loads far from the buckling threshold, where the response remains approximately linear elastic \cite{Feher2024}. We introduce three simple parameterized cut patterns and perform a parametric study to explore the relationship between the cut locations and the structural rigidity under a concentrated vertical load. Accordingly, three cases are considered: Case 1 and 2 investigate how we can improve the rigidity of an arch made of a non-perforated, rectangular sheet by additional cuts in the case of symmetric and asymmetric loading, respectively; Case 3 investigates whether and how it is possible to improve the rigidity of a kirigami arch having an existing perforation. We compared the numerical results to experiments to support our findings. 

Section \ref{sec:theory} presents the mechanical model and solution approach, Section \ref{sec:results} discusses the results for all three cases, and Section \ref{sec:conclusion} summarizes the findings.

\section{Theoretical background}
\label{sec:theory}
\subsection{Mechanical model}
\begin{figure}[htbp]
    \centering
    \includegraphics[keepaspectratio,width=\linewidth]{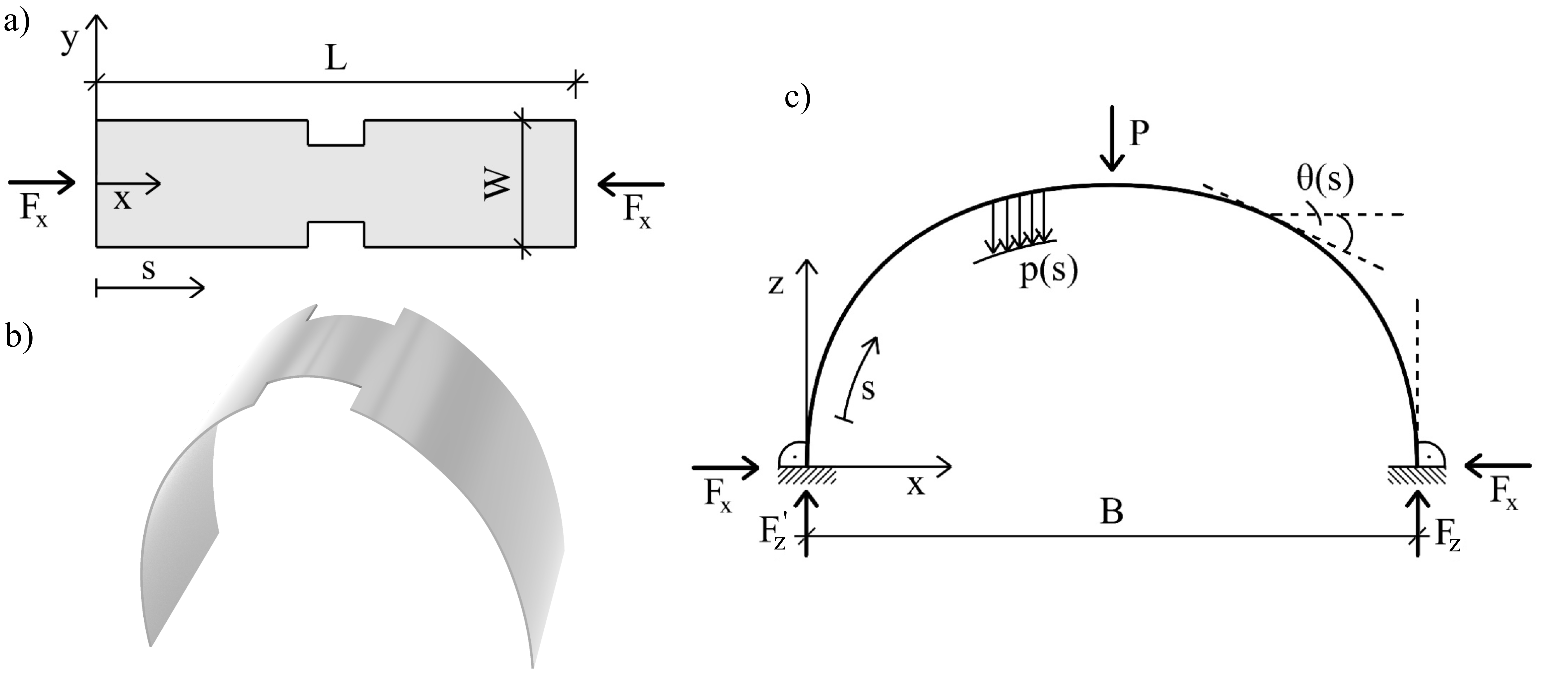}
    \caption {a) Initial configuration, top view of the initially flat sheet. The curved shape is enforced by two horizontal forces $F_x$ acting at its ends. b) Visualization of the resulting three-dimensional arch. c) Mechanical model of the bending-active arch. The ends of the sheet are clamped and at a vertical position and fixed at a distance $B$. The arch carries its self-weight $p(s)$ and a concentrated load $P$. The vertical loads are balanced by vertical support reactions $F'_z$ and $F_z$.}
    \label{fig:model}
\end{figure}

We consider an initially flat, rectangular sheet of length $L$, width $W$, and thickness $t$ decorated by a cut pattern having reflection symmetry to the longitudinal axis $x$ (Fig. \ref{fig:model}a). Two horizontal forces are applied at the ends of the sheet, and the flat geometry is transformed into a bending-active arch (Fig. \ref{fig:model}b). The ends of the sheet are fixed at a distance $B<L$, with the tangent constrained to be vertical at the supports. In this deformed configuration, internal forces arise in the structure even without any external loads, since the initial geometry is flat. The sheet carries its self-weight, a non-uniformly distributed load that depends on the cut pattern. In the following, we refer to this configuration of the structure as the unloaded state. Then, we apply a concentrated load on the arch, resulting in the loaded state (Fig. \ref{fig:model}c).  

Such bending-active arches can be modeled using a nonlinear beam model, the so-called elastica equation \cite{howell2009applied,Liu2020,Zhang2022,Feher2024}, where the changes along the thickness of the sheet are neglected and the deformation of the sheet is considered to be constant in the $y$ direction. The sheet is parameterized by its arc-length $s \in [0,L]$. The moment equilibrium equation under non-uniformly distributed vertical load, and a concentrated vertical load acting at $s_P$ reads as:
\begin{equation} \begin{split} \label{eq:elastica0}
E\frac{\mathrm{d}}{\mathrm{d}s} \left[ I_{\mathrm{eff}}(s) \frac{\mathrm{d}\theta(s)}{\mathrm{d}s} \right] - F_z \cos{\theta(s)} + F_x \sin{\theta(s)} \\ + \cos{\theta(s)} \int_{0}^s p(s) \,\mathrm{d}s + P \cos{\theta(s)} H(s-s_P) = 0,
\end{split}
\end{equation}
where $\theta (s)$ is the tangent angle of the beam, $E$ is Young's modulus, $I_{\mathrm{eff}}(s)$ is the effective moment of inertia, $H$ is the Heavyside function, $p(s)$ is the distributed load, $P$ is the concentrated vertical load, $F_x$ is the horizontal support reaction, $F_z'$ and $F_z$ are the vertical support reactions.

The perforations are taken into account in the distributed load and the effective moment of inertia. The width of the cross-section changes along the longitudinal axis. The total width of the solid part of the sheet is denoted by $w(s)$. Consequently, the self-weight is
\begin{equation}
p(s) = w(s) t \rho,
\end{equation}
where $\rho$ is the volume density of the material. Applying the technique described in \cite{Feher2024}, we define the effective width $w_{\mathrm{eff}}(s)$ using a parameter $0\leq\alpha\leq\pi$ to take into account the changed the stress distribution around the cuts. Regions around the cuts at an angle $\alpha$ from the $y$ direction are considered to be nearly stress-free. In case of $\alpha = 0$, $w_{\mathrm{eff}}(s) = w(s)$, while $\alpha = \pi/2$ corresponds to a constant $w_{\mathrm{eff}}(s) = \min(w(s))$ function. Here, we considered $\alpha=\pi/4$. Accordingly, the moment of inertia is
\begin{equation}
I_{\mathrm{eff}}(s) = \frac{w_{\mathrm{eff}}(s) t^3}{12}.
\end{equation} 
The global porosity of the sheet is $\Phi = (A_0 - A)/A_0$, where the area of the bounding rectangle (i.e., the non-perforated sheet) is $A_0 = LW$ and $A = \int_0^L w(s) \, \mathrm{d}s$ is the area of the solid part. 

The relationship between the tangent angle and the position of the points is:
\begin{equation} 
\frac{\mathrm{d} z}{\mathrm{d}s} = \sin{\theta},
\end{equation}
\begin{equation} 
\frac{\mathrm{d} x}{\mathrm{d}s} = \cos{\theta}.
\end{equation}
There are four boundary conditions
\begin{equation} \label{eq:bc1}
\theta(0) = \pi/2,
\end{equation}
\begin{equation} \label{eq:bc2}
\theta(L) = - \pi/2,
\end{equation}
\begin{equation} \label{eq:bc3}
x(L) = \int_0^{L} \cos(\theta(s)) \, \mathrm{ds} = B,
\end{equation}
\begin{equation} \label{eq:bc4}
z(L) = \int_0^{L} \sin(\theta(s)) \, \mathrm{ds} = 0.
\end{equation}
prescribing the tangent and the position of the sheet at the supports, respectively. The $\Delta L$ axial shortening of the sheet is neglected. 

\subsection{Numerical solution}
Let
\begin{equation}
Q = \int_0^L p(s) 
\end{equation}
denote the amplitude of the distributed load, and 
\begin{equation}
q(s) = \frac{ \int_0^s p(s) }{Q}
\end{equation}
its normalized cumulative distribution.
Accordingly, the equilibrium equation \eqref{eq:elastica0} becomes:
\begin{equation} \label{eq:elastica}
E\frac{\mathrm{d}}{\mathrm{d}s} \left[ I_{\mathrm{eff}}(s) \frac{\mathrm{d}\theta(s)}{\mathrm{d}s} \right] + F_x \sin{\theta(s)} + \left( Q q(s) + P H(s-s_P) - F_z \right) \cos{\theta(s)}  = 0.
\end{equation}

The system has six parameters: $F_z,F_x,Q,P,x(L),z(L)$. We solved it using the pde2path \cite{Uecker_Wetzel_Rademacher_2014,dohnal2014pde2path,uecker2021numerical} and Chebfun \cite{driscoll2014chebfun} toolboxes in Matlab. To ensure convergence, the problem is solved in three continuation steps, with $F_x,F_z$ always free parameters. First, we set the distance between the supports without external loads ($P=Q=0$) starting from a stable shape corresponding to $F_x=0.001$ and run the continuation in the $x(L)$ parameter until eq. \eqref{eq:bc3} is satisfied. Then the self-weight is applied by numerical continuation in $Q$. These first two steps result in the $z_o(s)$ unloaded shape of the structure. Finally, $P$ is applied in the third numerical continuation step to obtain the $z(s)$ loaded shape. 

The rigidity of the structure is defined from the unloaded ($Q\neq0, P = 0$) and loaded ($Q\neq0, P\neq0$) configurations as 
\begin{equation} \label{eq:rigidity}
K = \frac{P}{z(s_P)-z_o(s_P)}.
\end{equation}

\section{Results and discussion}
\label{sec:results}
We investigated the relationship between cut geometry and structural rigidity by designing three simple, parameterized cut patterns. Using the methodology outlined in Section \ref{sec:theory}, we applied a $P = 1N$ concentrated load and computed both the rigidity and height of the resulting arches as functions of the pattern parameters. Then we selected some patterns for experimental analysis and compared them to the numerical results. For the loading range considered, the arches behaved approximately linearly elastic, and the applied load remained well below the snap‐through threshold in all cases. This was confirmed by numerical continuation, which showed that all computed configurations stayed within the stable range.

The experiments were conducted using polyethylene terephthalate (PET) sheets of thickness $t=0.5\,\mathrm{mm}$. The Young modulus of the sheet was approximated by $E=2800\,\mathrm{N/mm^2}$ \citep{polymer} and the density $\rho = 1.33\, \mathrm{g/cm^3}$ was provided by the manufacturer. The patterns were cut out using a laser cutter. The bounding rectangle of the patterns was 360 mm x 90 mm with $10\, \mathrm{mm}$ for the clamps, leaving a length $L = 340\, \mathrm{mm}$ and width $W = 90\, \mathrm{mm}$ for the structure. The self-weight of this sheet is non-negligible, as the length exceeds the elasto-gravitational length \citep{vella2009statics}: 
\begin{equation}
	l_g = \Bigl( \frac{EWt^3/12}{\rho g t} \Bigr) ^{1/3} \leq  L.
\end{equation}

\subsection{Case 1: Symmetric load} \label{sec:case1}

\begin{figure}[htbp]
    \centering
    \includegraphics[keepaspectratio,width=0.4\linewidth]{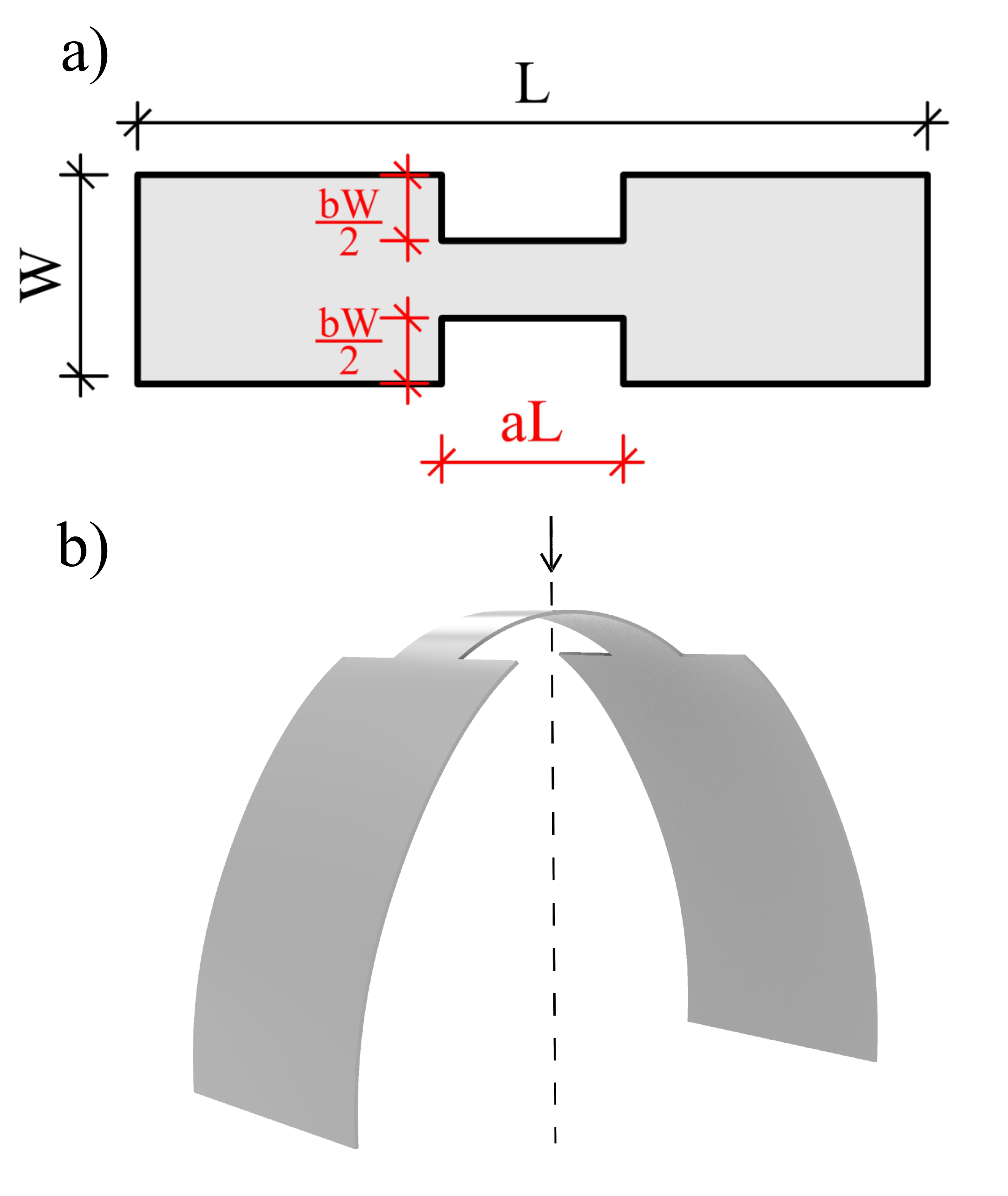}
    \caption {a) Cut pattern of Case 1. There are two symmetrically placed rectangular cuts at the center of the edges of the sheet characterized by parameters $a$ and $b$. In case $a=b=0$, the sheet is non-perforated. b) Three-dimensional illustration of the shape and the load.}
    \label{fig:pattern1}
\end{figure}

In this section, we investigate the effect of rectangular cuts on an initially non-perforated sheet under concentrated load acting in the middle of the structure ($s_P = 0.5 L$). The cut pattern is defined by two parameters, $a$ and $b$, within the ranges $0.05 \leq a \leq 0.9$ and $0.05 \leq b \leq 0.9$ values (Fig. \ref{fig:pattern1}). The cuts are positioned at the center of the edges of the sheet. 

For each parameter combination, the height ($h = z(0.5L)$) and the structural rigidity ($K$) were computed for three different support distances ($B = 0.25L,\, 0.5L,\, 0.75L$).
The baseline height ($h_o = z_o^{a=b=0}(0.5L)$) and rigidity ($K_o = K^{a=b=0}$) values of the non-perforated sheet ($a=b=0$) are summarized in Table \ref{tab:case1} along with the $h_{max},K_{max}$ maximum values for the investigated parameter ranges.

\begin{table}[htbp]
\centering
\begin{tabular}{ c|c|c|c }
B & $\mathbf{0.25L}$ & $\mathbf{0.5L}$ & $\mathbf{0.75L}$ \\
\hline
$h_o$ [mm] & 142.1 & 125.5 & 86.7 \\ 
$h_{max}$ [mm] & \red{157.1} & \red{136.7} & \red{92.12}  \\
\hline
$K_o$ & 0.266 & 0.213 & 0.129 \\ 
$K_{max}$ & \red{0.612} & \red{0.296} & \blue{0.128} \\
\hline
\end{tabular}
\caption{Calculated rigidity and height values for Case 1. Red and blue indicate an increase or decrease in rigidity compared to the baseline, respectively.}
\label{tab:case1}
\end{table}

\begin{figure}[htbp]
    \centering
    \includegraphics[keepaspectratio,width=\linewidth]{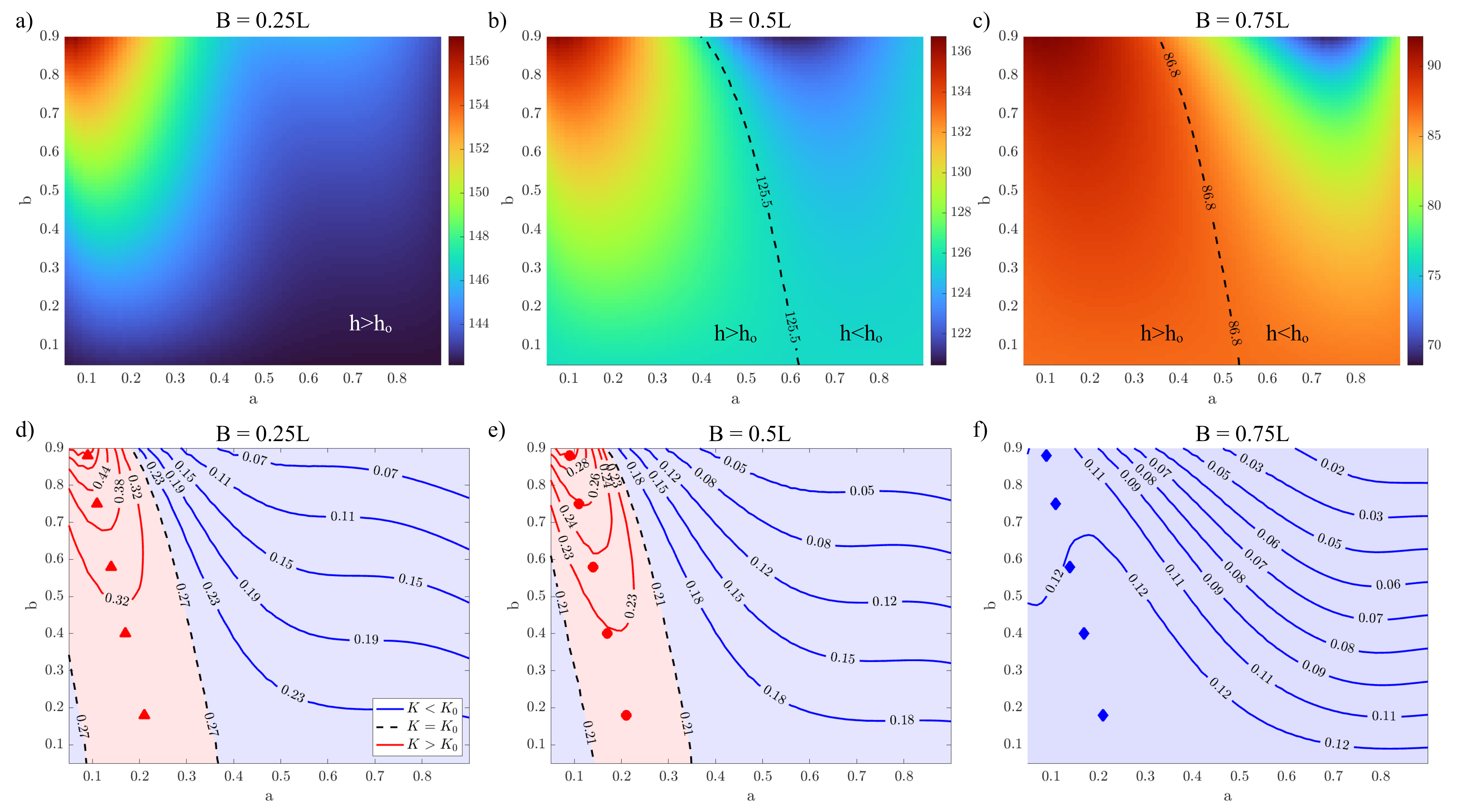}
    \caption {Numerical results of Case 1. The black dashed line represents the baseline values from Table \ref{tab:case1}. a)-c) Height $h$ as a function of $a$ and $b$ for each support distance. d)-f) Contour plots of rigidity $K$ as a function of $a$ and $b$ for each support distance. Red and blue indicate increased or decreased rigidity relative to the baseline, respectively. Marked points indicate patterns selected for experimental testing. }
    \label{fig:case1_rh}
\end{figure}

Fig. \ref{fig:case1_rh}) shows the calculated height and rigidity values in the function of the pattern parameters. The results show that introducing cuts can increase the unloaded height of the structure, though this does not necessarily imply higher rigidity. For $B=0.25L$, all investigated cut patterns resulted in a higher unloaded shape than the non-perforated sheet (Fig.~\ref{fig:case1_rh}a), yet rigidity increased only within a limited region of the $(a,b)$ parameter space (Fig.~\ref{fig:case1_rh}d). Interestingly, $a,b$ values that are too small or too large, do not increase the rigidity. Increasing the distance of the supports ($B=0.5L$) resulted in the decrease of the range of favorable parameter combinations: only smaller cut widths resulted in an increased height (Fig. \ref{fig:case1_rh}b)) and the range of $a,b$ values leading to increased rigidity shrank (Fig. \ref{fig:case1_rh}e)). Finally, if the supports are too far from each other ($B=0.75L$), no parameter combination resulted in rigidity increase (Fig. \ref{fig:case1_rh}f)), although the height increase was still possible (Fig. \ref{fig:case1_rh}c)). 

\begin{figure}[htbp]
    \centering
    \includegraphics[keepaspectratio,width=\linewidth]{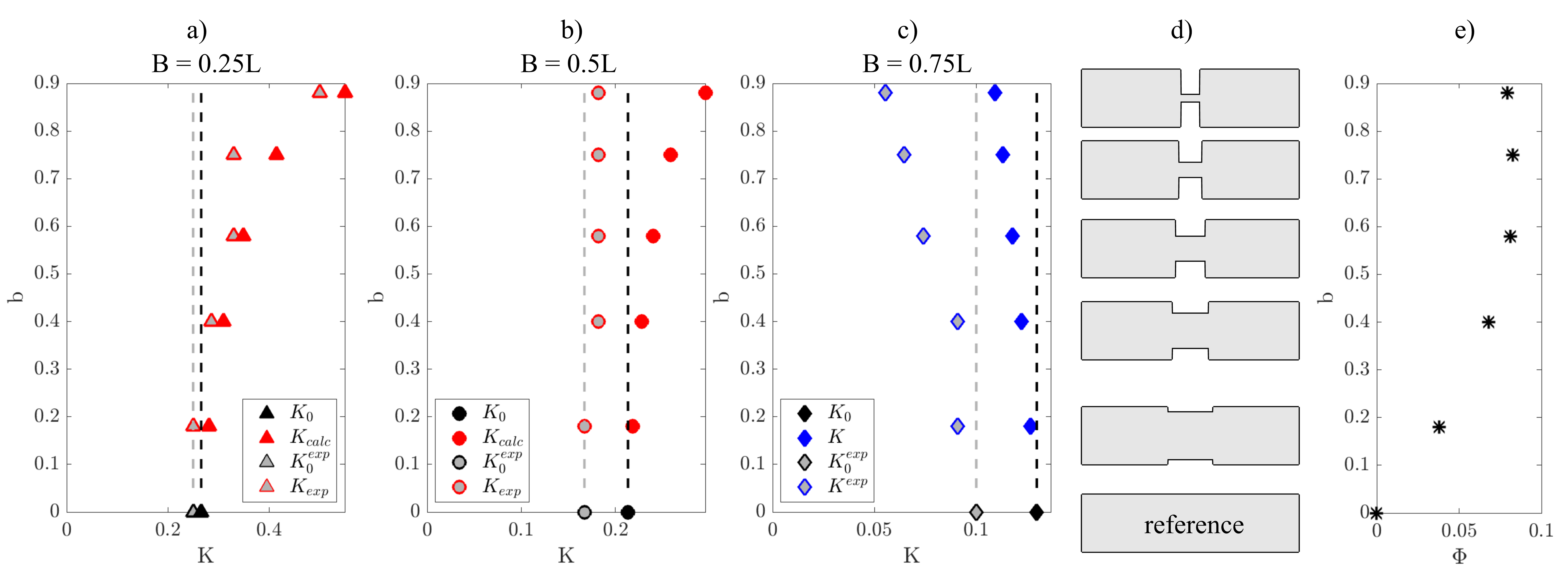}
    \caption {a)-c) Experimental and numerical rigidity results for the selected cut patterns for each support distance. Experimental results are distinguished by grey fills. Black and dashed lines indicate baseline values (black for calculations, grey for experiments). Red and blue denote increased or decreased rigidity compared to the baseline. d) Visualization of the selected cut patterns aligned with the markers on the diagrams. e) Global porosity $\Phi$ as a function of the pattern.}
    \label{fig:case1_ce}
\end{figure}

Five cut patterns, in addition to the reference (non-perforated) sheet, were selected from the diagrams in Fig. \ref{fig:case1_rh}d–f) for experimental validation. The comparison between numerical predictions and experiments is shown in Fig. \ref{fig:case1_ce}a)-c), along with visualizations of the tested patterns (Fig. \ref{fig:case1_ce}d) and the corresponding global porosity (Fig. \ref{fig:case1_ce}e). Overall, the experiments and the calculations show good agreement: rigidity increased for $B=0.25L$ and $B=0.5L$ for all the chosen patterns (Fig. \ref{fig:case1_ce}a)-b)) but decreased for $B=0.75L$ (Fig. \ref{fig:case1_ce}c)). Discrepancies between the values could be attributed to the uncertainties and precision limitations of the measurements. A comparison between Figs.~\ref{fig:case1_ce}a–c) and Fig. \ref{fig:case1_ce}e) indicates a non-monotonic relationship between global porosity $\Phi$ and rigidity.

\begin{figure}[htbp]
    \centering
    \includegraphics[keepaspectratio,width=0.5\linewidth]{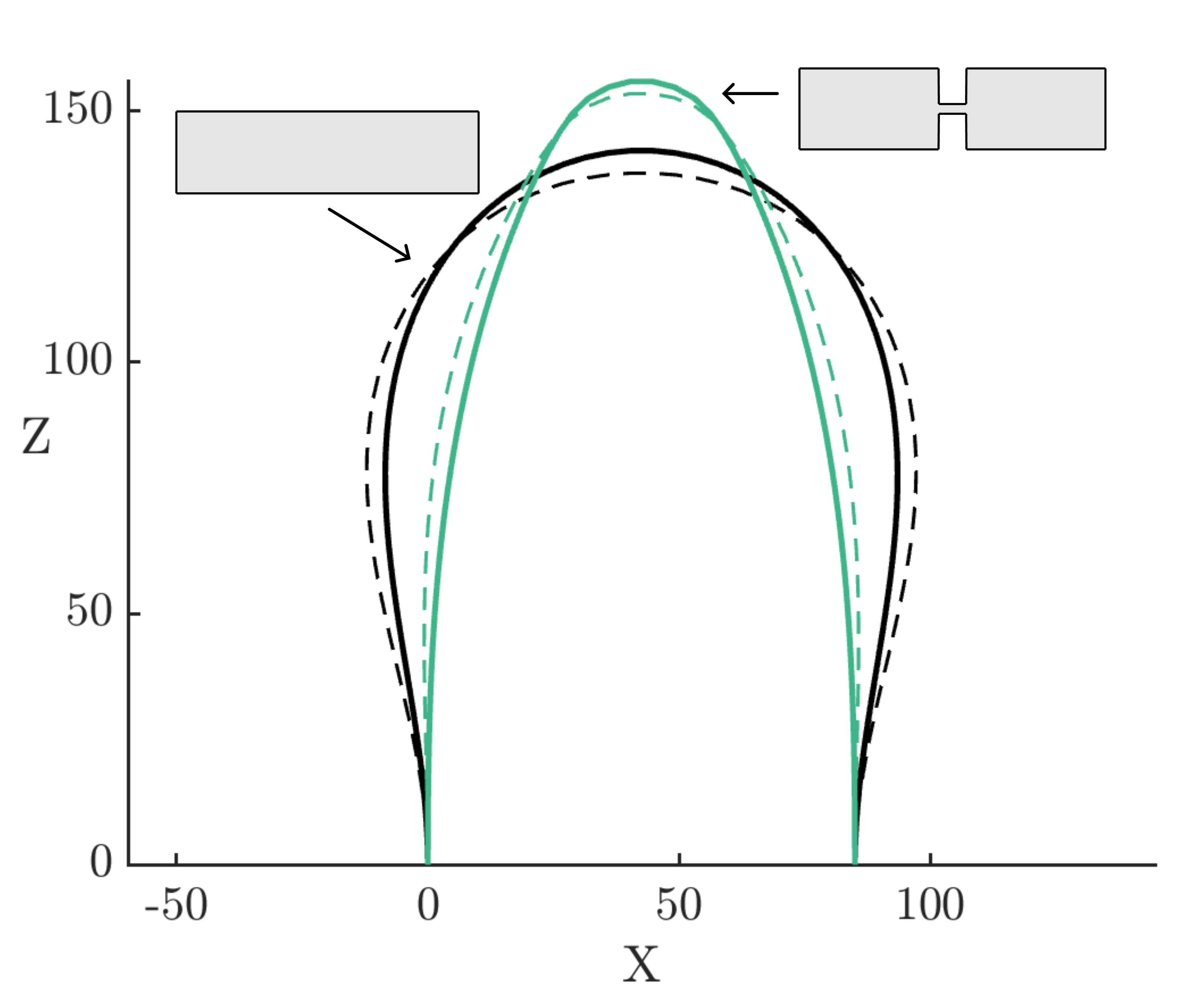}
    \caption {Calculated shape of the reference sheet and a pattern with $a = 0.09, b=0.88$ for $B=0.25L$. Solid and dashed lines denote unloaded and loaded configurations, respectively.}
    \label{fig:case1_shape}
\end{figure}

\begin{figure}[htbp]
    \centering
    \includegraphics[keepaspectratio,width=\linewidth]{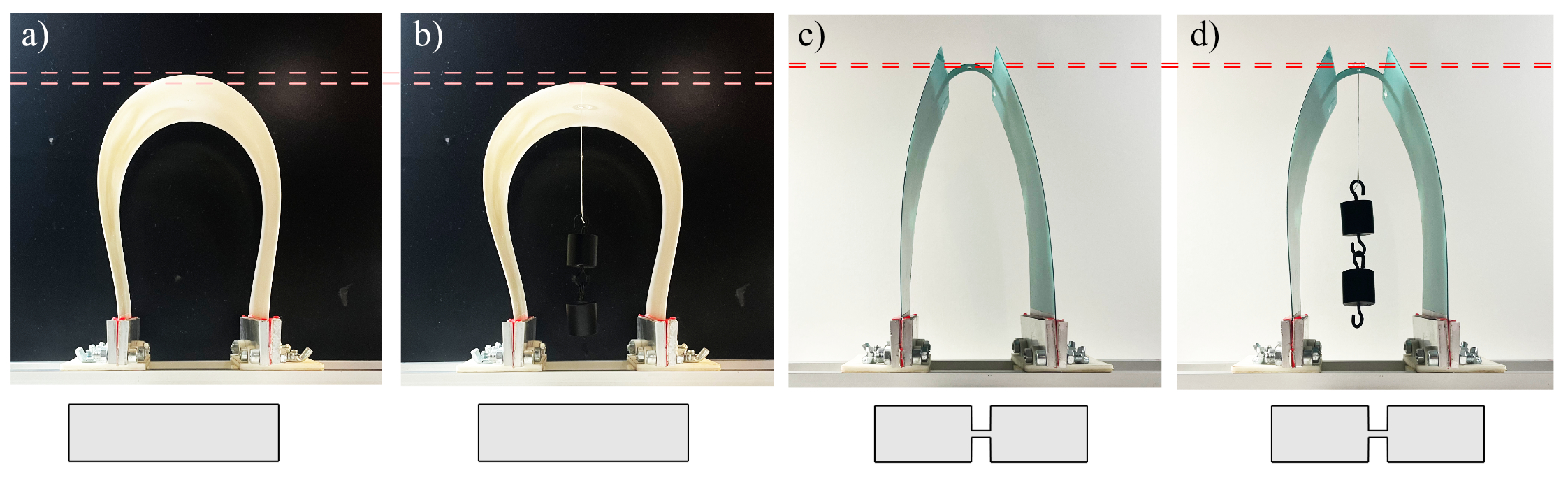}
    \caption {Experimental photographs of Case 1 for $B=0.25L$. The corresponding cut patterns are shown under the photographs. Pink dashed lines mark the loaded and unloaded heights for reference. a)-b) Unloaded and loaded shape of the reference pattern, respectively. c)-d) Unloaded and loaded shape of the cut pattern $a = 0.09, b=0.88$, respectively.}
    \label{fig:case1_exp}
\end{figure}

For $B=0.25L$, the most favorable of the selected cut patterns ($a=0.09, b=0.88$) exhibited more than twice the rigidity of the reference sheet. Fig. \ref{fig:case1_shape} shows the computed shapes, and Fig. \ref{fig:case1_exp} presents corresponding experimental photographs. The cut pattern resulted in a higher unloaded shape, and it deflected less both in the experiments and the calculations compared to the reference.

These results show two findings. Firstly, under certain conditions, it is possible, to increase the rigidity of the structure by introducing appropriately placed cuts, revealing a rigidity paradox: increasing the global porosity, i.e., weakening the cross-section of the structure, could also increase the rigidity. Secondly, while cuts may result in a higher unloaded shape, increased height does not necessarily correlate with increased rigidity.

\subsection{Case 2: Asymmetric load} \label{sec:case2}
\begin{figure}[htbp]
    \centering
    \includegraphics[keepaspectratio,width=0.4\linewidth]{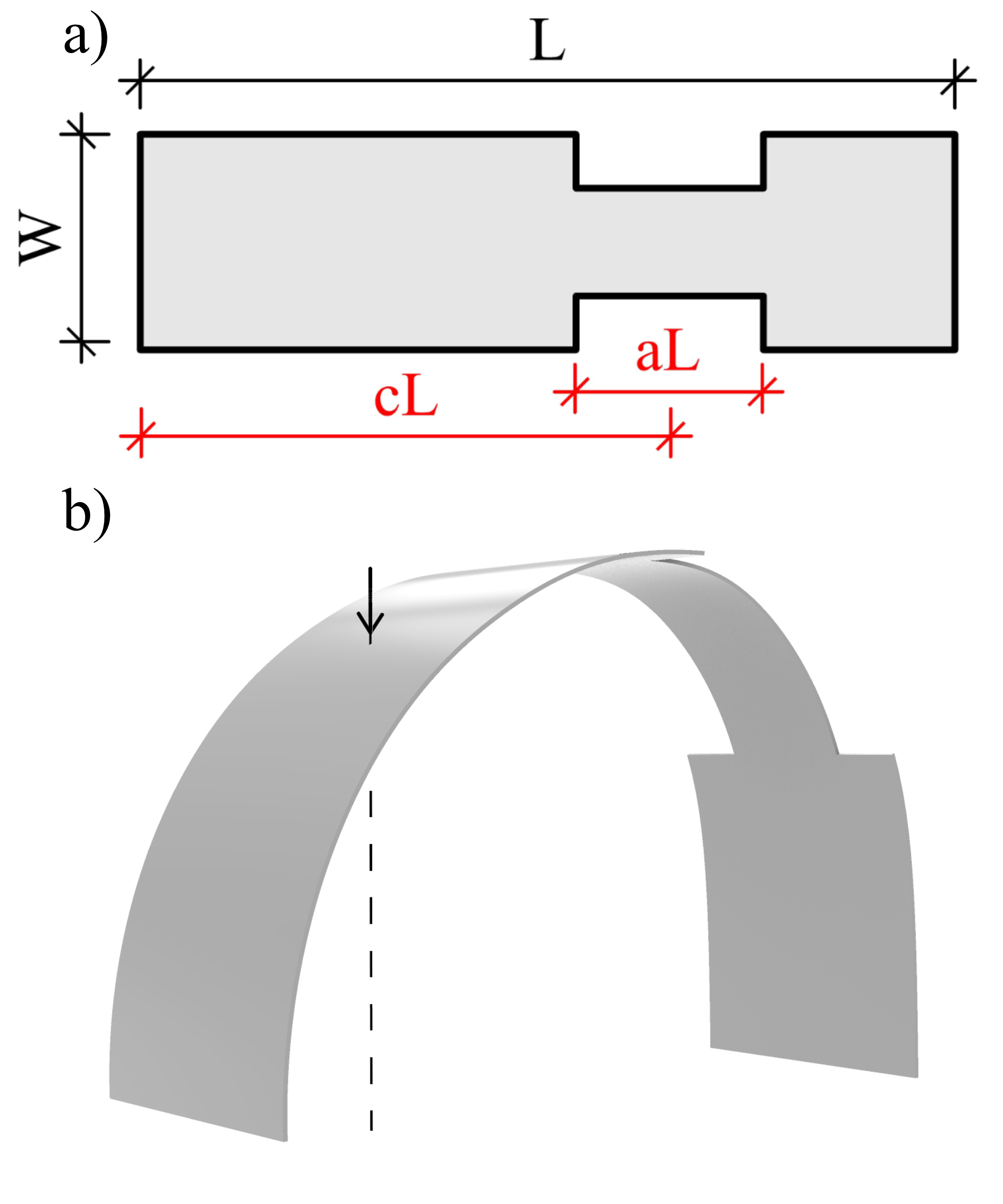}
    \caption {a) Cut pattern of Case 2 characterized by parameters $a$ and $c$, and symmetric only about the longitudinal axis and. b) Three-dimensional illustration of the shape and the load acting at $s_P = 0.33L$.}
    \label{fig:pattern2}
\end{figure}

The next case investigates the connection between the load position and the cut location on the sheet. The load is applied at the left third of the sheet, specifically at $s_P = 0.33L$. The cut pattern parameterized by $a$ and $c$ is illustrated in Fig. \ref{fig:pattern2}. The height of the cuts is fixed at $bW/2$ with $b = 0.5$. We conducted a parameter study in $0.05 \leq a \leq 0.9$ and $0.05 \leq c \leq 0.9$ for three different support distances ($B=0.25L,0.5L,0.75L$). The cuts are constrained so that they do not cut through the entire sheet, i.e., only a limited range of $a,c$ pairs satisfying $cL - aL/2 > 0$ are considered. Baseline height $h_o$ and rigidity $K_o$ values of the non-perforated sheet ($a=b=c=0$) were calculated for reference and listed in Table \ref{tab:case2} along with the $h_{max},K_{max}$ maximum values for the investigated parameter ranges.

\begin{table}[htbp]
\centering
\begin{tabular}{ c|c|c|c }
B & $\mathbf{0.25L}$ & $\mathbf{0.5L}$ & $\mathbf{0.75L}$ \\
\hline
$h_o$ [mm] & 113.3 & 105.8 & 78.8 \\ 
$h_{max}$ [mm] & \red{114.4} & \red{109.1} & \red{81.86}  \\
\hline
$K_o$ & 2.131 & 0.371 & 0.155 \\ 
$K_{max}$ & \red{69.025} & \red{2.110} & \red{0.382} \\
\hline
\end{tabular}
\caption{Calculated rigidity and height values for Case 2. Red and blue colors correspond to an increased or decreased rigidity, respectively.}
\label{tab:case2}
\end{table}

\begin{figure}[htbp]
    \centering
    \includegraphics[keepaspectratio,width=\linewidth]{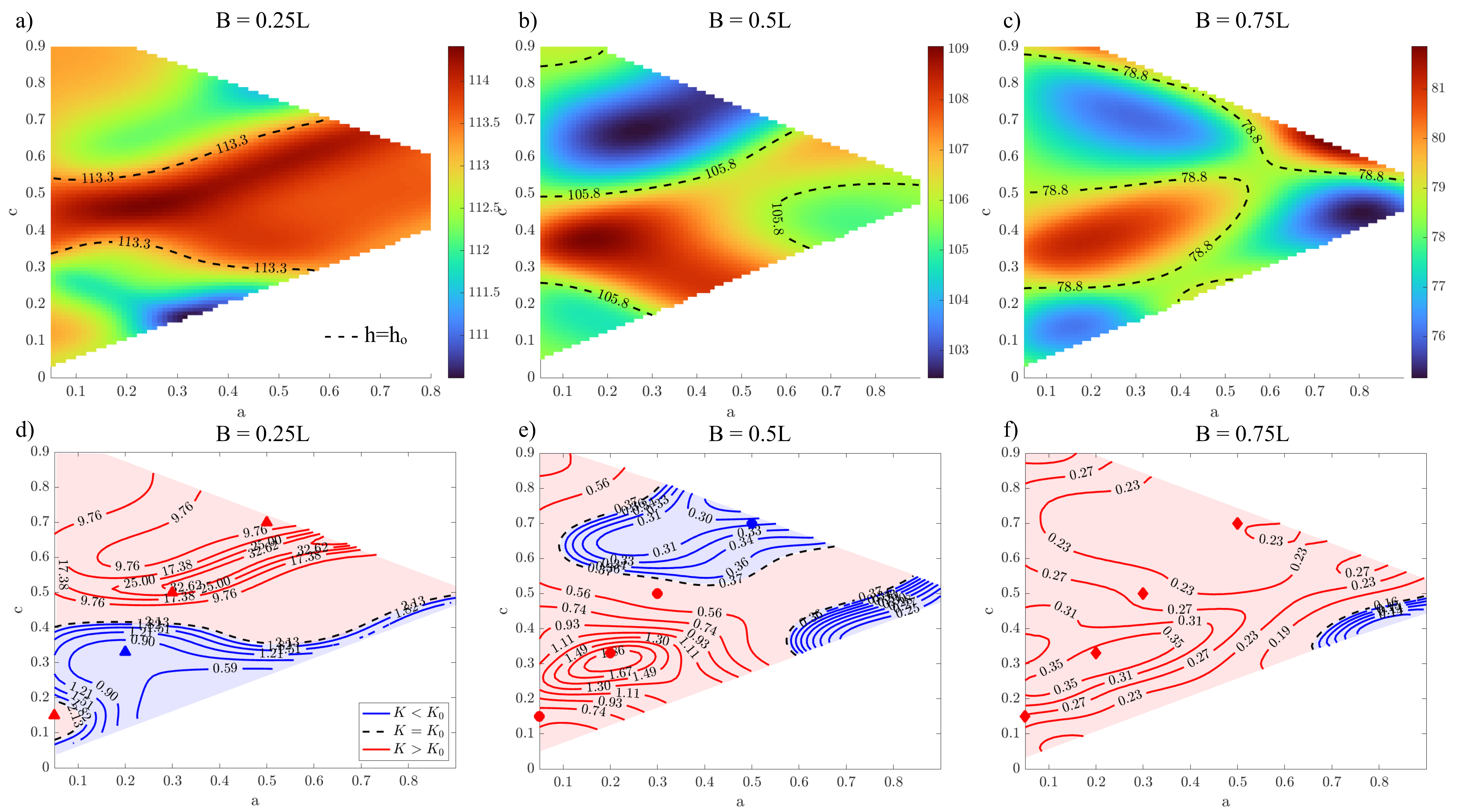}
    \caption{Numerical results of Case 2. The black dashed line represents the baseline values from Table \ref{tab:case1}. a)-c) Height $h$ as a function of $a$ and $c$ for each support distance. d)-f) Contour plots of rigidity $K$ as a function of $a$ and $c$ for each support distance. Red and blue indicate increased or decreased rigidity relative to the baseline, respectively. Marked points indicate patterns selected for experimental testing.}
    \label{fig:case2_rh}
\end{figure}

Figure \ref{fig:case2_rh} displays the calculated height and rigidity as functions of the parameters $a$ and $c$ for each support distance. Interestingly, in Figure \ref{fig:case2_rh}d)-f), the range of favorable parameters (highlighted by red) is non-convex in all three cases. Rigidity and height can be increased for all support distances, but the optimal cut location (i.e., leading to the largest rigidity or height) depends highly on the support distance. For $B=0.5L$ and $B=0.75L$, the most favorable location of the cut is near the acting point of the load (Fig. \ref{fig:case2_rh}e)-f)), as cuts allow for higher curvature and increase the height of the structure (Fig. \ref{fig:case2_rh}b)-c)) at their location, consistent with observations from Case 1. Conversely, placing the cuts near the acting point of the load is unfavorable for $B=0.25L$ (Fig. \ref{fig:case2_rh}d)), and the range of favorable parameters is split into two parts.

\begin{figure}[htbp]
    \centering
    \includegraphics[keepaspectratio,width=\linewidth]{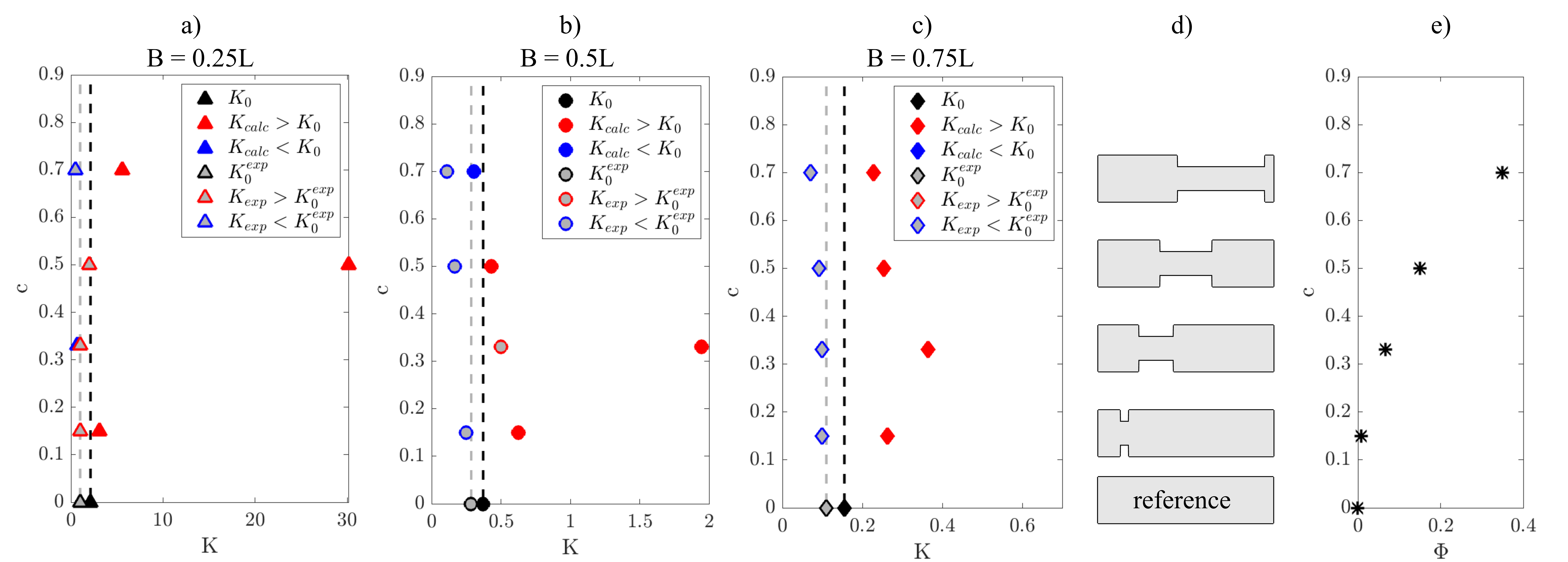}
     \caption {a)-c) Experimental and numerical rigidity results of Case 2 for the selected cut patterns for each support distance. Experimental results are distinguished by grey fills. Experimental values are shown with grey fills. Black and dashed lines indicate baseline values (black for calculations, grey for experiments). Red and blue denote increased or decreased rigidity compared to baseline. d) Visualization of the selected cut patterns aligned with the markers on the diagrams. e) Global porosity $\Phi$ as a function of the parameter pattern.}
    \label{fig:case2_ce}
\end{figure}

Four cut patterns, indicated in Fig. \ref{fig:case2_rh}d)-f), were selected for experimental analysis. Fig. \ref{fig:case2_ce} shows the comparison of the numerical (color fills) and experimental (grey fills) rigidity values for these selected patterns (red and blue markers) and the reference pattern (black). For $B=0.25L$ (Fig. \ref{fig:case2_ce}a)), the pattern with $c = 0.5$ performed best both in the experiments and calculations. The calculated deflection of this pattern was very close to zero (0.05 mm), and it was over the precision of our measurements, resulting in a significant difference between the calculations and measurements. For $B=0.5L$ (Fig. \ref{fig:case2_ce}b)), the $c=0.33$ pattern performed the best in both the experiments and calculations. Finally, for $B=0.75L$ (Fig. \ref{fig:case2_ce}c)), contrary to the calculations, we were not able to improve the rigidity by cuts in the experiments. This discrepancy may stem from uncertainties in defining the effective moment of inertia, laser-cutting inaccuracies, and sensitivity to load asymmetry. Figure \ref{fig:case2_ce}e) shows the global porosity aligned with the corresponding cut patterns of Fig. \ref{fig:case2_ce}d) for reference. 

\begin{figure}[htbp]
    \centering
    \includegraphics[keepaspectratio,width=0.5\linewidth]{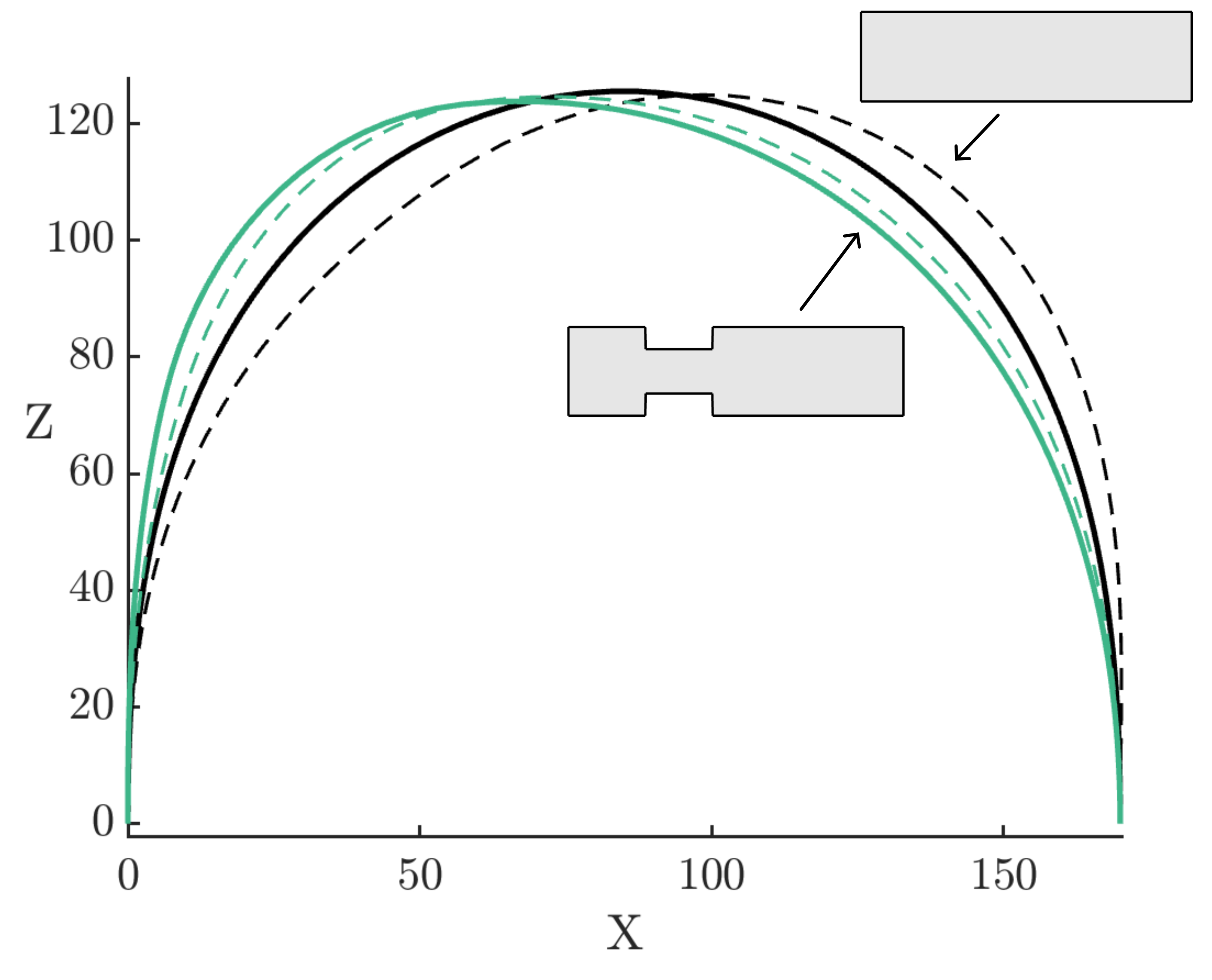}
    \caption {Calculated shape of the reference sheet and a pattern with $a=0.2,c=0.33$ for $B=0.5L$ for Case 2. Solid and dashed lines denote unloaded and loaded configurations, respectively.}
    \label{fig:case2_shape}
\end{figure}

\begin{figure}[htbp]
    \centering
    \includegraphics[keepaspectratio,width=\linewidth]{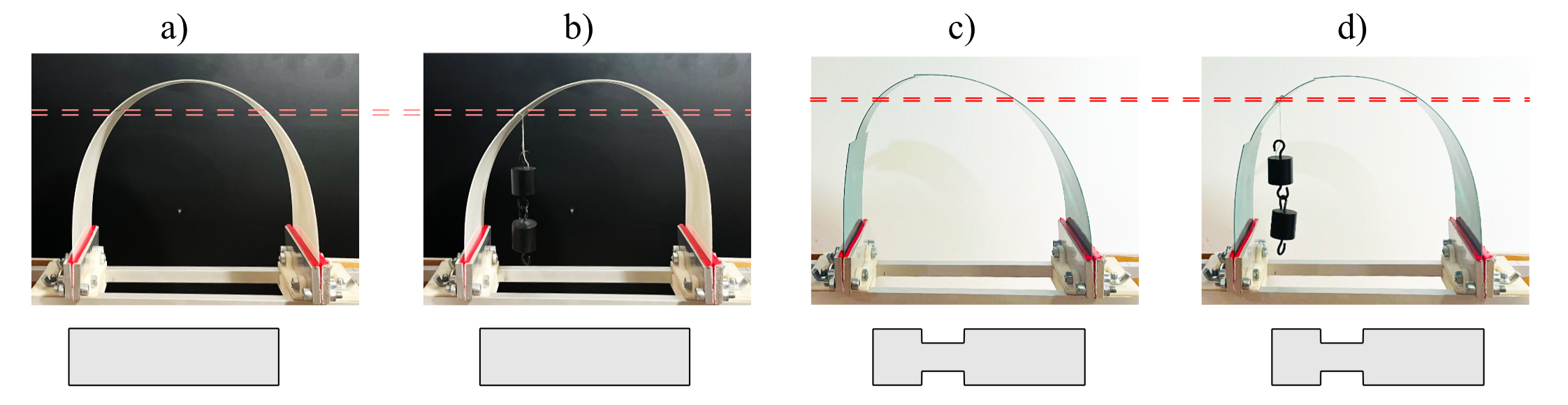}
    \caption {Experimental photographs of Case 2 for $B=0.5L$. The corresponding patterns are shown under the photographs. Pink dashed lines mark the height of the structures for reference. a)-b) Unloaded and loaded shape of the reference pattern, respectively. c)-d) Unloaded and loaded shape of the cut pattern $a=0.2,c=0.33$, respectively.}
    \label{fig:case2_exp}
\end{figure}

The calculated shape of the most favorable pattern ($a=0.2,c=0.33$) from the selected patterns for $B=0.5L$ is shown in Fig. \ref{fig:case2_shape} along with the reference. While the reference pattern exhibits increased asymmetry under the load, the chosen pattern reduces asymmetry, resulting in increased rigidity. Experimental photos (Fig. \ref{fig:case2_exp}) illustrated these findings: the initially symmetric reference pattern becomes asymmetric under the load (Fig. \ref{fig:case2_exp}a)-b)), whereas the initially asymmetric cut pattern gains back symmetry after load application (Fig. \ref{fig:case2_exp}c)-d)).

\begin{figure}[htbp]
    \centering
    \includegraphics[keepaspectratio,width=\linewidth]{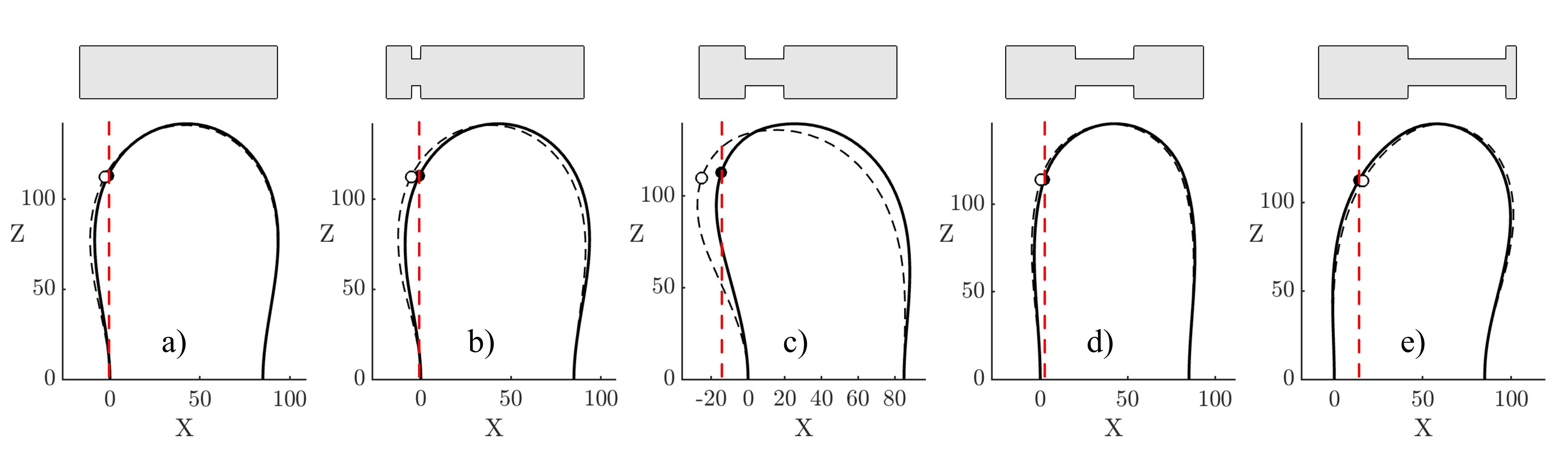}
    \caption {a)-e) Calculated shapes of Case 2 for $B=0.25L$, the solid black line is the unloaded shape, the dashed black line is the loaded shape. Filled and void dots correspond to the $s_P=0.33L$ in the unloaded and loaded states, respectively. Red dashed line shows the line of action of the load for the unloaded shape.}
    \label{fig:case2_explanation}
\end{figure}

The case $B=0.25L$ differs notably from the others in two aspects: (1) placing the cut around the acting point of the load does not improve rigidity, (2) the optimal cut yields rigidity that substantially exceeds the reference. Fig. \ref{fig:case2_explanation} shows the calculated shapes for all of the selected patterns. The line of of action (red dashed lines) reveals that $c=0.33$ (Fig. \ref{fig:case2_explanation}c)) increases asymmetry and the arm of the load, thereby the deflection is larger. On the other hand, $c=0.5$ (Fig. \ref{fig:case2_explanation}d)) keeps the line of action close to the support, and it also straightens the sheet segment between the left support and the acting point, which is structurally advantageous. 

In conclusion, rigidity can still be improved under asymmetric loading by carefully positioned cuts; however, the structural performance is heavily affected by the position of the load and the symmetry parameters of the resulting shape. 

\subsection{Case 3: Effect of the initial pattern} \label{sec:case3}
\begin{figure}[htbp]
    \centering
    \includegraphics[keepaspectratio,width=0.4\linewidth]{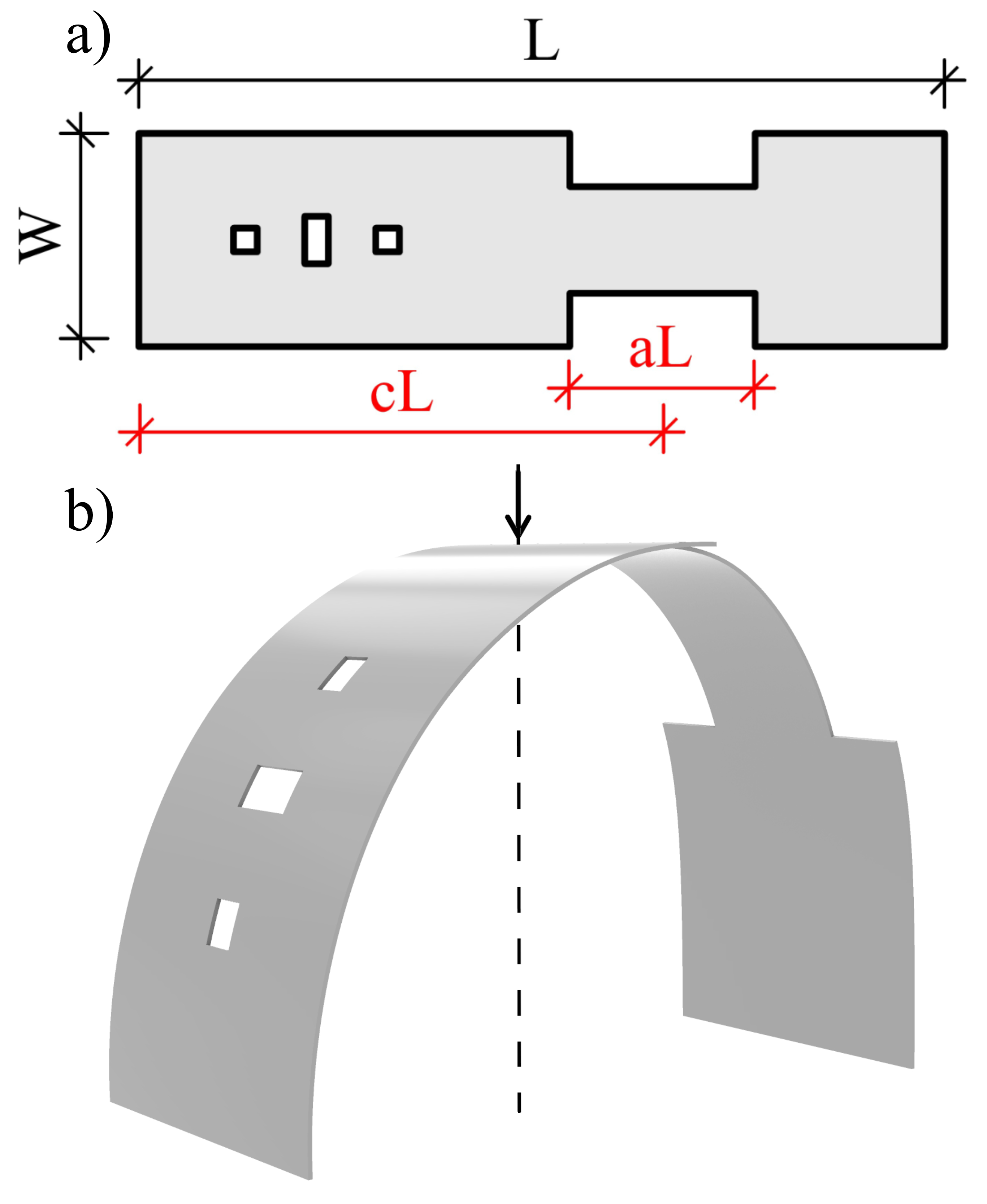}
    \caption {a) Cut pattern of Case 3. The initial perforation consists of three rectangular holes measuring 2.5x2.5mm, 2.5x5mm, and 2.5x2.5mm, spaced 5 mm apart. They are centered in the $y$-direction, with the first hole positioned 10 mm from the left edge of the sheet. The additional cuts are defined by parameters $a$ and $c$. b) Three-dimensional illustration of the shape and the load acting at $s_P = 0.5L$.}
    \label{fig:pattern3}
\end{figure}

This subsection aims to determine whether additional cuts can compensate for the unfavorable existing pattern and improve rigidity. In this case, the cut pattern features a non-symmetric initial perforation (Fig. \ref{fig:pattern3}) with two side cuts parameterized by $0.05 \leq a \leq 0.8$ and $0.05 \leq c \leq 0.95$, subject to $cL - aL/2 > 0$. The height of the cuts is $bW/2$ with $b = 0.5$. The load acts at the symmetry axis at $s_P = 0.5L$. Based on the results from Subsections \ref{sec:case1} and \ref{sec:case2}, the initial cut pattern is unfavorable for the symmetric load position. We considered three support distances as in the previous cases ($B=0.25L,0.5L,0.75L$).

The $h_o,K_o$ baseline values for the reference pattern ($a=b=c=0$ and the initial perforation) are summarized in Table \ref{tab:case3} along with the maximum $h_{max},K_{max}$ values of the calculated ranges. Note that the rigidity of the reference pattern is slightly worse than the rigidity of a non-perforated sheet carrying the same load (Table \ref{tab:case1}). Correspondingly, the best parameter combinations can restore the height and the rigidity.

\begin{figure}[htbp]
    \centering
    \includegraphics[keepaspectratio,width=\linewidth]{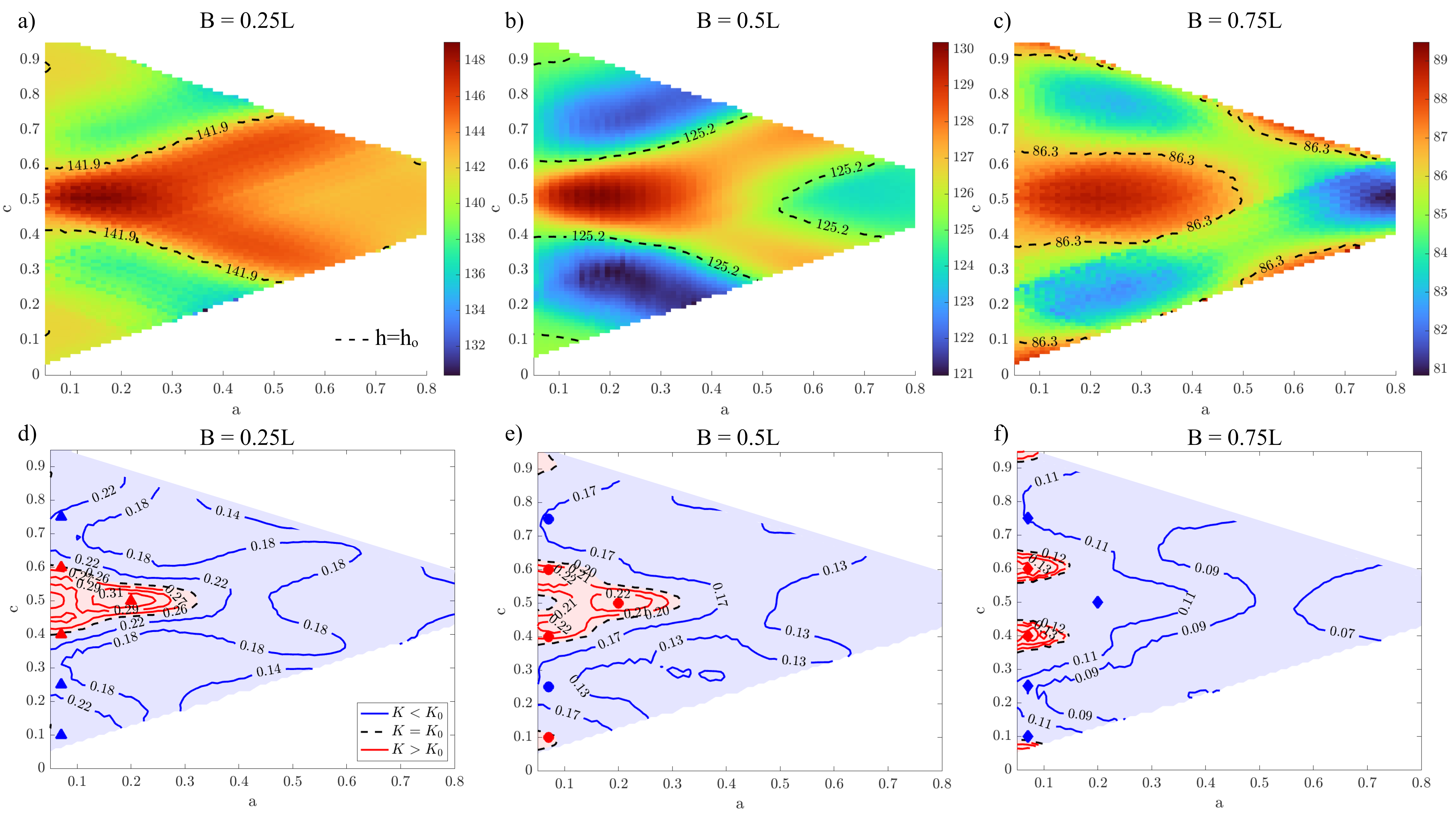}
    \caption {Numerical results of Case 3. The black dashed line represents the baseline values from Table \ref{tab:case3}. a)-c) Height $h$ as a function of $a$ and $c$ for each support distance. d)-f) Contour plots of rigidity $K$ as a function of $a$ and $c$ for each support distance. Red and blue indicate increased or decreased rigidity relative to the baseline, respectively. Marked points indicate patterns selected for experimental testing. }
    \label{fig:case3_rh}
\end{figure}

\begin{table}[htbp]
\centering
\begin{tabular}{ |c|c|c|c| }
\hline
B & $\mathbf{0.25L}$ & $\mathbf{0.5L}$ & $\mathbf{0.75L}$ \\
\hline
$h_o$ [mm] & 141.9 & 125.2 & 86.3 \\ 
$h_{max}$ [mm] & \red{149.0} & \red{130.4} & \red{89.5}  \\
\hline
$K_o$ & 0.256 & 0.204 & 0.124 \\ 
$K_{max}$ & \red{0.335} & \red{0.235} & \red{0.130} \\
\hline
\end{tabular}
\caption{Calculated rigidity and height values for Case 3. Red and blue indicate an increase or decrease in rigidity compared to the baseline, respectively.}
\label{tab:case3}
\end{table}
Figure \ref{fig:case3_rh} presents the calculated height and rigidity values in terms of parameters $a$ and $c$ for each support distance. Interestingly, the favorable range of parameters is split into multiple distinct regions and they suggest that asymmetric cut positions could also be beneficial. Moreover, none of the parameter combinations result in a symmetric cut pattern, but the diagrams are very close to being symmetric. For $B=0.25L$, the optimal cut location is near the acting point of the load (Fig. \ref{fig:case3_rh}d)). For $B=0.5L$ and $B=0.75L$, multiple regions exhibit increased rigidity, including areas near the supports (Fig. \ref{fig:case3_rh}e)-f)). Additionally, for $B=0.5L$, a symmetric cut ($c=0.5$) only increases rigidity if it is sufficiently large ($0.09 \leq a$), otherwise, the rigidity decreases. Finally, for $B=0.75L$, symmetric cuts do not improve rigidity regardless of size, consistent with observations in Figure \ref{fig:case1_rh}f). 

\begin{figure}[htbp]
    \centering
    \includegraphics[keepaspectratio,width=\linewidth]{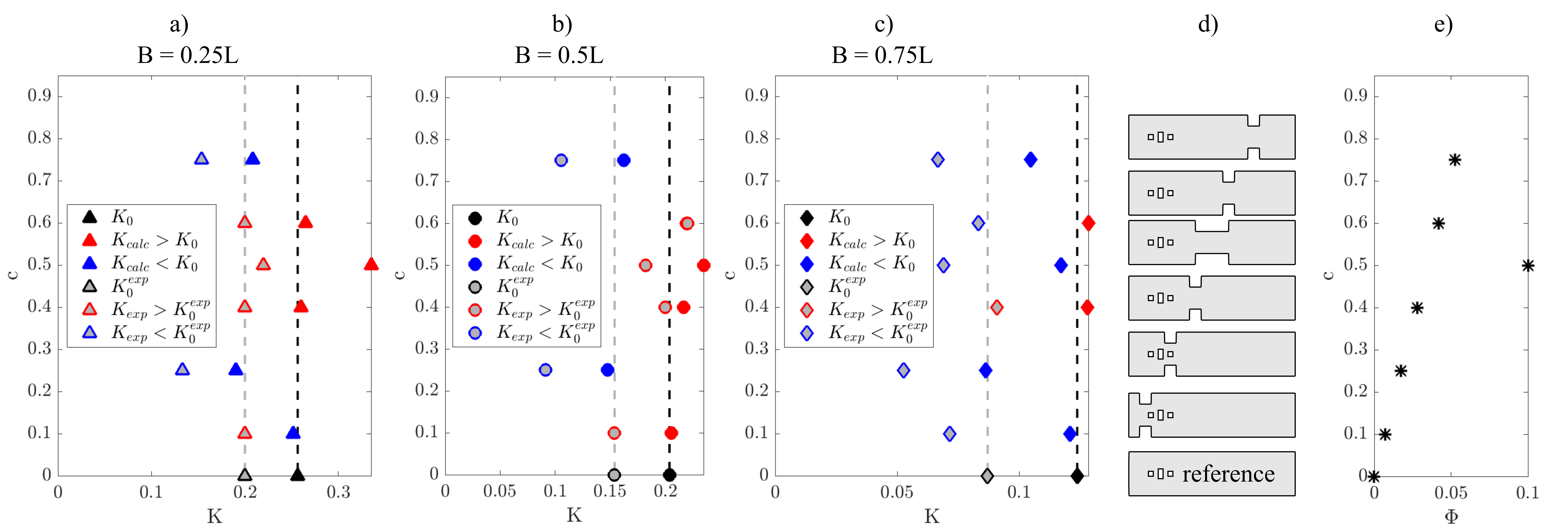}
    \caption {a)-c) Experimental and numerical rigidity results of Case 3 for the selected cut patterns for each support distance. Experimental results are distinguished by grey fills. Experimental values are shown with grey fills. Black and dashed lines indicate baseline values (black for calculations, grey for experiments). Red and blue denote increased or decreased rigidity compared to baseline. d) Visualization of the selected cut patterns aligned with the markers on the diagrams. e) Global porosity $\Phi$ as a function of the parameter pattern.}
    \label{fig:case3_ce}
\end{figure}

Six cut patterns representing various regions of the rigidity diagrams were selected for experimental validation (Fig. \ref{fig:case3_rh}d)-f)). Figure \ref{fig:case3_ce} compares the calculated and measured rigidity values. $c=0.5$ coincides with the load position and it is the favorable cut for $B=0.25L$ and $B=0.5L$, but it decreases rigidity for $B=0.75L$ (Fig. \ref{fig:case3_ce}a)-c)). Experiments confirm that asymmetric cut positions could also increase the rigidity (red markers for $c\neq0.5$). The global porosity diagram (Fig. \ref{fig:case3_ce}e)) reveals no simple correlation between global porosity and rigidity. 

\begin{figure}[htbp]
    \centering
    \includegraphics[keepaspectratio,width=0.5\linewidth]{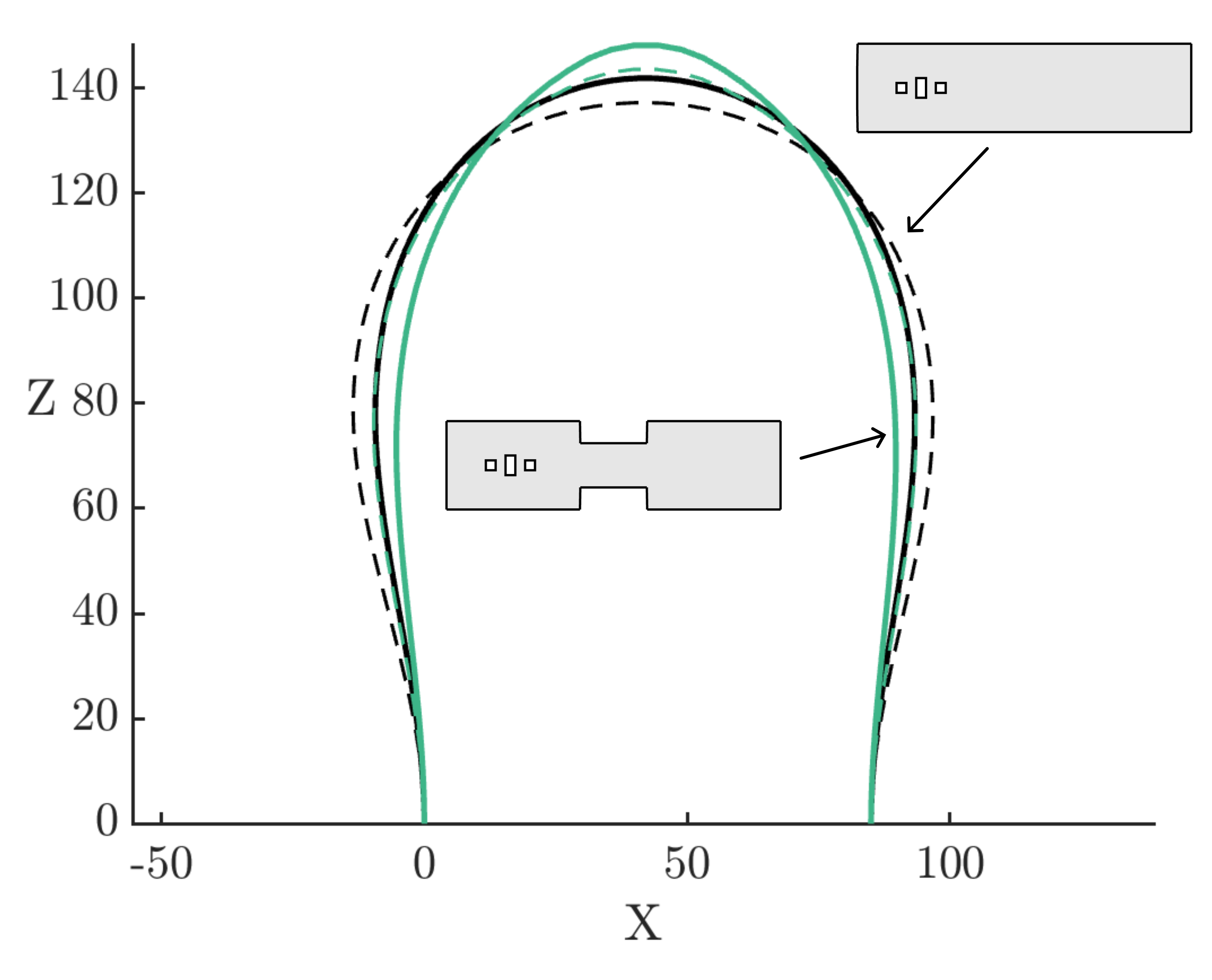}
    \caption{Calculated shape of the reference sheet and a pattern with $a=0.2, c=0.5$ for $B=0.25L$ and Case 3. Solid and dashed lines mark the unloaded and loaded shapes, respectively.}
    \label{fig:case3_shape}
\end{figure}

\begin{figure}[htbp]
    \centering
    \includegraphics[keepaspectratio,width=\linewidth]{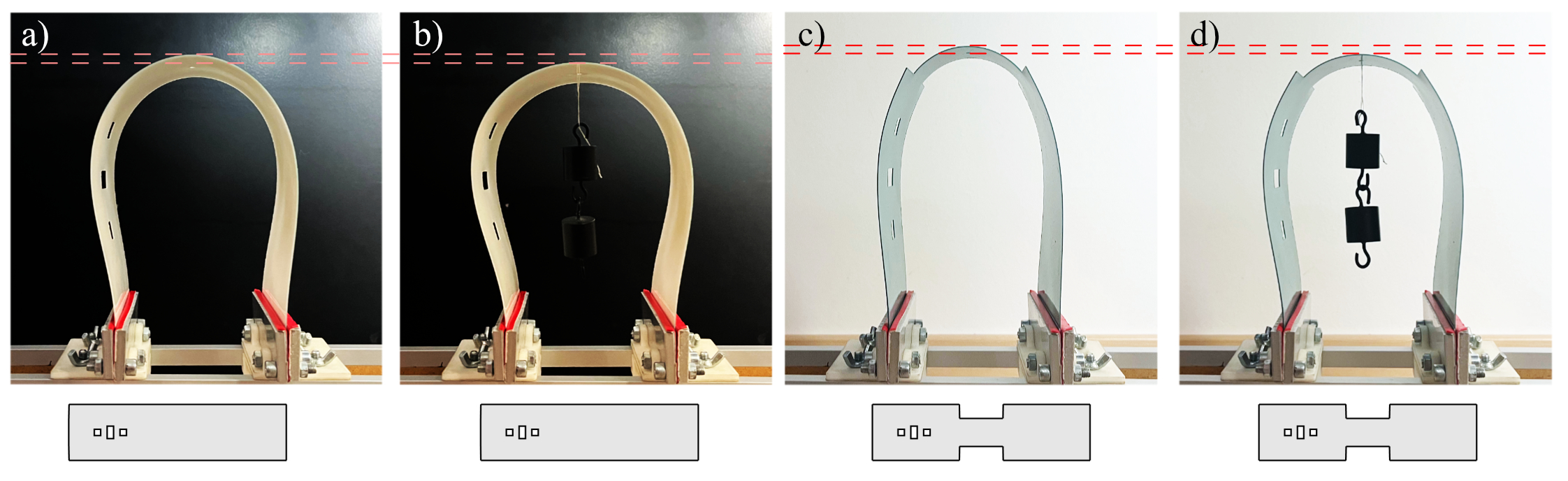}
    \caption {Experimental photographs of Case 3 for $B=0.25L$. The corresponding cut patterns are shown under the photographs. Pink dashed lines mark the loaded and unloaded heights for reference. a)-b) Unloaded and loaded shape of the reference pattern, respectively. c)-d) Unloaded and loaded shape of the cut pattern $a=0.2, c=0.5$}
    \label{fig:case3_exp}
\end{figure}

Figure \ref{fig:case3_shape} shows the calculated shapes of the reference pattern and the most favorable from the investigated patterns in case $B=0.25L$ and Fig. \ref{fig:case3_exp} shows the corresponding experimental photographs. The centered cuts allow for larger curvatures in the middle region, which produces taller and structurally preferable shapes.

Case 3 demonstrates that the mechanical performance of perforated kirigami arches can be improved by additional cuts. It was also shown that for the investigated pattern, the additional cuts could restore the original rigidity of the structure. These findings also highlight that the optimal cut placement depends on multiple factors, and even under symmetric loading conditions, asymmetric cuts may outperform symmetric ones.

\section{Conclusion}
\label{sec:conclusion}
Contrary to intuition, introducing cuts on a kirigami arch does not always weaken the structure. Our findings reveal a rigidity paradox: depending on the load position and support distance, the cuts can increase not only the height but also the rigidity of the structure. By locally weakening cross-sections, the sheet can deform into shapes that better resist the applied load. This study aimed to determine whether, and to what extent, the rigidity of a kirigami arch can be increased through cuts under various load positions and support distances, and to examine the influence of pre-existing cut patterns.

We introduced three parametric cut patterns to systematically explore the interplay between the cut geometry, the support distances, rigidity, and height. First, we demonstrated that rigidity gains are possible for initially non-perforated sheet both under symmetric (Sec. \ref{sec:case1}) and asymmetric (Sec. \ref{sec:case2}) loading conditions. Next, we analyzed the impact of an existing cut pattern (Sec. \ref{sec:case3}). Our results show that the relationship between the global porosity and the rigidity is non-monotonic. Additionally, the rigidity gain achievable by the cuts decreases as the support distance increases. 

Selected patterns were validated experimentally, with measured rigidity improvements generally smaller but consistent with numerical predictions. In some cases, both experiments and simulations achieved more than double the original rigidity (e.g., Fig. \ref{fig:case1_ce}a, Fig. \ref{fig:case1_exp}). The largest discrepancy between experiments and simulations occurred under asymmetric loading (Sec. \ref{sec:case2}), highlighting the need for further investigation.

This work focused on programmable rigidity in kirigami arches. Future research could extend these concepts to simultaneously control shape and mechanical performance in more complex kirigami structures such as domes or curved surfaces, enabling designs that meet both aesthetic and structural criteria.

\section*{Acknowledgments}
The author would like to thank Tomas Dohnal, Hannes Uecker, and Alexander Meiners for the fruitful discussions and their help with pde2path. We acknowledge the Digital Government Development and Project Management Ltd. for awarding us access to the Komondor HPC facility based in Hungary. This research was supported by the NKFIH Hungarian Research Fund Grants 143175, the NKFIH 24 Advanced 149429, and TKP2021-NVA BME. This research was conducted with the support of the ERASMUS+ Programme of the European Union.



 \bibliographystyle{elsarticle-num} 
 \bibliography{cas-refs}





\end{document}